\documentclass[letterpaper,12pt,oneside]{article}

\usepackage{endfloat}
\usepackage{amsfonts}
\usepackage{mychicago}
\usepackage{subfigure}
\usepackage{genetics_manu_style}

\usepackage[dvips]{color}
\usepackage{graphicx}
\usepackage{amsmath}
\usepackage{times}
\usepackage[scanall]{psfrag}
\newcommand{\ratet}{\mbox{$\alpha_t$}}
\newcommand{\rateinf}{\mbox{$\alpha$}}
\newcommand{\Npop}{\mbox{$N$}}

\newcommand{\mavg}{\mbox{$\langle m \rangle_{T}$}}
\newcommand{\msqavg}{\mbox{$\langle m^2 \rangle_{T}$}}
\newcommand{\mavgsmall}{\mbox{$\langle m \rangle_{T, o}$}}
\newcommand{\msqavgsmall}{\mbox{$\langle m^2 \rangle_{T, o}$}}
\newcommand{\mavginf}{\mbox{$\langle m \rangle_{T, \infty}$}}
\newcommand{\savg}{\mbox{$\langle s \rangle_{T}$}}

\newcommand{\ddg}{\mbox{$\Delta\Delta G$}}

\newcommand{\dgfmin}{\mbox{$\Delta G_{f}^{\rm{min}}$}}
\newcommand{\dgf}{\mbox{$\Delta G_{f}$}}
\newcommand{\dgfmut}{\mbox{$\Delta G_{f}^{\rm{mut}}$}}
\newcommand{\dgfwt}{\mbox{$\Delta G_{f}^{\rm{wt}}$}}
\newcommand{\dgfextra}{\mbox{$\Delta G_{f}^{\rm{extra}}$}}
\newcommand{\conf}{\mbox{$\mathcal{C}$}}
\newcommand{\conft}{\mbox{$\mathcal{C}_t$}}

\newcommand{\p}{\mbox{$\mathbf{p}$}}
\newcommand{\x}{\mbox{$\mathbf{x}$}}
\newcommand{\avec}{\mbox{$\mathbf{a}$}}
\newcommand{\rvec}{\mbox{$\mathbf{r}$}}
\newcommand{\lvec}{\mbox{$\mathbf{l}$}}
\newcommand{\W}{\mbox{$\mathbf{W}$}}
\newcommand{\Q}{\mbox{$\mathbf{Q}$}}
\newcommand{\A}{\mbox{$\mathbf{A}$}}
\newcommand{\Hm}{\mbox{$\mathbf{H}$}}
\newcommand{\Wij}{\mbox{$W_{ij}$}}
\newcommand{\Hji}{\mbox{$H_{ji}$}}

\newcommand{\V}{\mbox{$\mathbf{V}$}}
\newcommand{\I}{\mbox{$\mathbf{I}$}}

\newcommand{\Vii}{\mbox{$V_{ii}$}}
\newcommand{\nui}{\mbox{$\nu_i$}}
\newcommand{\nuj}{\mbox{$\nu_j$}}

\newcommand{\pavg}{\mbox{$\mathbf{p_o}$}}
\newcommand{\xinf}{\mbox{$\mathbf{x_{\infty}}$}}
\newcommand{\ainf}{\mbox{$\mathbf{a_{\infty}}$}}
\newcommand{\nuavg}{\mbox{$\langle \nu \rangle$}}
\newcommand{\nusmall}{\mbox{$\langle \nu \rangle_{o}$}}
\newcommand{\nut}{\mbox{$\langle \nu \rangle_{t}$}}
\newcommand{\nuinf}{\mbox{$\langle \nu \rangle_{\infty}$}}
\newcommand{\iod}{\mbox{$R_{T}$}}
\newcommand{\iodsmall}{\mbox{$R_{T,o}$}}
\newcommand{\iodinf}{\mbox{$R_{T,\infty}$}}
\newcommand{\pmt}{\mbox{$\mathbf{p}\left(m, t\right)$}}
\newcommand{\xmt}{\mbox{$\mathbf{x}\left(m, t\right)$}}

\newcommand{\pmtplus}{\mbox{$\mathbf{p}\left(m, t + 1\right)$}}
\newcommand{\pmtminus}{\mbox{$\mathbf{p}\left(m, t - 1\right)$}}
\newcommand{\pmminust}{\mbox{$\mathbf{p}\left(m - 1, t\right)$}}
\newcommand{\xmtplus}{\mbox{$\mathbf{x}\left(m, t + 1\right)$}}

\newcommand{\xmminust}{\mbox{$\mathbf{x}\left(m - 1, t\right)$}}
\newcommand{\mkt}{\mbox{$\langle m^k \rangle_t$}}

\newcommand{\monet}{\mbox{$\langle m \rangle_{t}$}}
\newcommand{\monetplus}{\mbox{$\langle m \rangle_{t + 1}$}}
\newcommand{\mtwot}{\mbox{$\langle m^2 \rangle_{t}$}}
\newcommand{\mtwotplus}{\mbox{$\langle m^2 \rangle_{t + 1}$}}
\newcommand{\e}{\mbox{$\mathbf{e}$}}

\title{Thermodynamics Of Neutral Protein Evolution}
\author{Jesse D. Bloom,\thanks{Division of Chemistry and Chemical Engineering, California Institute of Technology, Pasadena, CA 91125} 
Alpan Raval,\thanks{Keck Graduate Institute of Applied Life Sciences and School of Mathematical Sciences, Claremont Graduate University, Claremont, CA 91711}
Claus O. Wilke\thanks{Section of Integrative Biology and Center for Computational Biology and Bioinformatics, University of Texas at Austin, Austin, TX 78712}}

\renewcommand{\CorrespondingAddress}{Division of Chemistry and Chemical Engineering \\ California Institute of Technology \\ Mail Code 210-41 \\ Pasadena, CA 91125 \\ (626) 354-2565 (ph.) \\ (626) 568-8743 (fax) \\ \texttt{jesse.bloom@gmail.com} \vfill}
\renewcommand{\RunningHead}{Thermodynamics Of Evolving Proteins}
\renewcommand{\CorrespondingAuthor}{Jesse D. Bloom}
\renewcommand{\KeyWords}{protein stability, mutational robustness, index of dispersion, molecular clock, lattice protein}

\begin{document}
\maketitle

\begin{abstract}
Naturally evolving proteins gradually accumulate mutations while continuing to fold to stable structures.  This process of neutral evolution is an important mode of genetic change, and forms the basis for the molecular clock. We present a mathematical theory that predicts the number of accumulated mutations, the index of dispersion, and the distribution of stabilities in an evolving protein population from knowledge of the stability effects (\ddg\ values) for single mutations. Our theory quantitatively describes how neutral evolution leads to marginally stable proteins, and provides formulae for calculating how fluctuations in stability can overdisperse the molecular clock. It also shows that the structural influences on the rate of sequence evolution observed in earlier simulations can be calculated using just the single-mutation \ddg\ values. We consider both the case when the product of the population size and mutation rate is small and the case when this product is large, and show that in the latter case the proteins evolve excess mutational robustness that is manifested by extra stability and an increase in the rate of sequence evolution.  All our theoretical predictions are confirmed by simulations with lattice proteins.  Our work provides a mathematical foundation for understanding how protein biophysics shapes the process of evolution.
\end{abstract}

\section{Introduction}
Proteins evolve largely through the slow accumulation of amino acid substitutions.  Over evolutionary time, this process of sequence divergence creates homologous proteins that differ at the majority of their residues, yet still fold to similar structures that often perform conserved biochemical functions~\cite{Lesk1980}.  The maintenance of structure and function during sequence divergence suggests that much of protein  evolution is neutral in the sense that observed sequence changes frequently do not alter a protein's ability to fold and adequately perform the biochemical function necessary to enable its host organism to survive.  This comparative evidence for neutrality in protein evolution has been corroborated by experimental studies showing that the mutations separating diverged sequences often have no effect other than modest and additive changes to stability~\cite{Serrano1993}, and that a large fraction of random mutations do not detectably alter a protein's structure or function~\cite{Shortle1985,Pakula1986,Loeb1989,Guo2004,Bloom2005,Bloom2006}. In this respect, it seems that protein evolution should be well described by  Kimura's neutral theory of evolution, which holds that most genetic change is due to the stochastic fixation of neutral mutations~\cite{Kimura1983}.  One of the key predictions  of the neutral theory is that assuming a constant mutation rate, the number of mutations separating two proteins should be proportional to the time since their divergence~\cite{Kimura1983}.  Indeed, the observation by \citeN{Zuckerkandl1965} that proteins are ``molecular clocks'' that accumulate mutations at a roughly constant rate has long been taken as one of the strongest pieces of evidence supporting the neutral theory~\cite{Ohta1971}.

However, mutations that are neutral with respect to a protein's capacity to perform its biological function often affect protein thermodynamics. The biological functions of most proteins depend on their  ability to fold to thermodynamically stable native structures~\cite{Anfinsen1973}. Yet natural proteins are typically only marginally stable, with free energies of folding (\dgf) between -5 and -15 kcal/mol~\cite{Fersht1999}.  Most random mutations to proteins are destabilizing~\cite{Godoy-Ruiz2004,Pakula1986,Matthews1993,Kumar2006}, and their effects on stability (measured as \ddg, the \dgf\ of the mutant protein minus the \dgf\ of the wildtype protein) are frequently of the same magnitude as a protein's net stability.  The impact of a mutation on a protein's function can therefore depend on the protein's stability: a moderately destabilizing mutation that is easily tolerated by a stable parent protein may completely disrupt the folding of a less stable parent.  This effect of protein stability on mutational tolerance has been verified by experiments demonstrating that more stable protein variants are markedly more robust to random mutations~\cite{Bloom2005,Bloom2006}.

The fact that mutations that are neutral with respect to direct selection for protein function can affect a protein's tolerance to subsequent mutations is not consistent with the simplest formulation of the neutral theory of evolution, which tends to assume that the fraction of mutations that is neutral remains constant in time.  \citeN{Kimura1987} himself recognized the possibility that the neutrality might change, and \citeN{Takahata1987} mathematically treated the consequences of a ``fluctuating neutral space''. In particular, Takahata showed that fluctuating neutrality could explain the observed overdispersion in the molecular clock~\cite{Cutler2000b}  (the tendency for the variance in the number of fixed mutations to exceed the expectation for the Poisson process predicted by the neutral theory) long considered troublesome for the neutral theory. However, further progress on this topic was stymied by the lack of a specific model for how or why protein neutrality might fluctuate.

More recently, researchers have preferred to describe neutral evolution using the concept of ``neutral networks,'' which are networks in the space of possible protein sequences in which each functional protein is linked to all other functional proteins that differ by only a single  mutation~\cite{Smith1970,Huynen1996,Govindarajan1997,vanNimwegen1999,Bornberg-Bauer1999,Tiana2000,Bastolla2002}.   A neutrally evolving protein population is then envisioned as moving on the neutral network, and the neutrality of the population may fluctuate if the nodes on the network differ in their connectivities.   A general theoretical treatment of evolution on neutral networks by \citeN{vanNimwegen1999} has shown that if the product of the population size and mutation rate is small then members of the population are equally likely to occupy any node, while if this product is large then the population will preferentially occupy highly connected nodes (see also \cite{Bornberg-Bauer1999,Taverna2002b,Xia2004b}).  Simulations with simplified lattice models of proteins have attempted to provide insight into the specific features of protein neutral networks. These simulations have shown that lattice protein neutral networks are centered  around highly connected nodes occupied by stable  proteins~\cite{Bornberg-Bauer1999,Xia2004b,Wingreen2004,Broglia1999}, a finding  consistent with the experimental observation~\cite{Bloom2005,Bloom2006} that stable proteins are more mutationally robust.   Lattice protein studies also suggest that protein structures differ in their ``designabilities'' (defined as the total number of sequences that fold into a structure), and that sequences that fold into more designable structures will neutrally evolve at a faster rate due to the increased size and connectivity of their neutral networks~\cite{Govindarajan1997,Li1996,England2003a,Chan2002,Wingreen2004}. Finally, simulations have demonstrated that fluctuations in neutrality as a protein population moves along its neutral network can lead to an overdispersion of the molecular clock~\cite{Bastolla2002}, as originally suggested by Takahata.  However, an extension of  these lattice protein simulations of evolution on neutral networks into a quantitative theory has been difficult because protein neutral networks are far too large to be computed for all but the simplest lattice models.

Here we present a mathematical treatment of neutral protein evolution that describes the evolutionary dynamics in terms of the \ddg\ values for single mutations, which are experimentally measurable.  Our treatment is based on the experimentally verified~\cite{Bloom2005,Bloom2006}  connection between protein stability and mutational robustness, as well as a few biophysically supported assumptions about \ddg\ values for random mutations.  By linking a protein's tolerance to mutations with stability, we are able to quantitatively describe neutral evolution without a full description of the neutral network.  We can then compute the average number of accumulated mutations, the average fraction of neutral mutations, the index of dispersion, and the  distribution of stabilities in a neutrally evolving population solely from knowledge of the \ddg\ values for single mutations.  In addition, we follow the formalism of \citeN{vanNimwegen1999} to calculate all four of these properties in the limit when the product of the population size and mutation rate is much less than one and in the limit when this product is much greater than one.  In demonstrating that these properties are different in these two limits, we show that  the rate of fixation of neutral mutations can vary with population size in violation of one of the standard predictions of Kimura's neutral theory~\cite{Kimura1987}.  Our work presents a unified view of neutral protein evolution that is grounded in measureable thermodynamic quantities.

\section*{Materials and Methods}
\subsection*{Lattice Protein Simulations}
We performed simulations with lattice proteins of $L = 20$ monomers of $20$ types corresponding to the natural amino acids. The proteins could occupy any of the 41,889,578 possible compact or non-compact conformations on a two-dimensional lattice.  The energy of a conformation \conf\ is the sum of the nonbonded nearest-neighbor interactions, $E\left(\conf\right) = \sum\limits_{i=1}^{L}\sum\limits_{j=1}^{i-2} C_{ij}\left(\conf\right) \times \epsilon\left(\mathcal{A}_i, \mathcal{A}_j\right)$, where $C_{ij}\left(\conf\right)$ is one if residues $i$ and $j$ are nearest neighbors in conformation \conf\ and zero otherwise, and $\epsilon\left(\mathcal{A}_i, \mathcal{A}_j\right)$ is the interaction energy between residue types $\mathcal{A}_i$ and $\mathcal{A}_j$, given by Table 5 of \citeN{Miyazawa1985}. We computed the stability of a conformation \conft\ as $\dgf\left(\conft\right) = E\left(\conft\right) + T \ln \left\{Q\left(T\right) - \exp\left[-E\left(\conft\right)/T\right]\right\},$ where $Q\left(T\right) = \sum\limits_{\left\{\conf_i\right\}} \exp \left[-E\left(\conf_i\right)/T\right]$ is the partition sum, made tractable by noting that there are only 910,972 unique contact sets.  All simulations were performed at a reduced temperature of $T = 1.0$

We used adaptive walks to find sequences that folded into each of the three arbitrarily chosen conformations shown in Fig. \ref{fig:lattice} with $\dgf \le 0$, and then neutrally evolved these sequences for $10^4$ generations with a population size of $N = 100$.  Our evolutionary algorithm was as follows: at each generation we randomly chose a protein that folded to the parental structure with $\dgf \le 0$ from the population and mutated each residue to some other randomly chosen residue with probability $5\times10^{-4}$, and continued doing this until we had filled the new population with proteins.  At the end of this equilibration evolution, we chose the most abundant sequence in the population as the starting point for further analysis and for the computation of the distribution of \ddg\ values for all 380 point mutations (sequences shown in Fig. \ref{fig:lattice}).  In principle, computing the distribution of \ddg\ values over all sequences in the population rather than just the most abundant one should give a more accurate representation of the true form of this distribution, and indeed we found that doing this slightly increased the accuracy of the predictions shown in Fig. \ref{fig:lattice}.  However, the resulting improvement in accuracy was small, since the approximate constancy of the \ddg\ distribution during neutral evolution (discussed below) means that the distribution computed over a single sequence is representative of that computed over all sequences in the population.  Therefore, we chose to compute the \ddg\ distribution over just the most abundant sequence since this choice more closely tracks what would be experimentally feasible with real proteins. (It is experimentally tractable to compute \ddg\ values for a single protein, but would be unmanageable to do so for all proteins in a natural population.)

To collect data for the case when the product $\Npop\mu$ of the population size \Npop\ and the per protein per generation mutation rate $\mu$ is $\ll 1$, we first equilibrated 1,000 replicates by evolving each of them with a population size of $N = 10$ and for 5,000 generations starting with a clonal population of the initial sequence described above.  The remainder of the evolutionary algorithm was as described above: the mutation rate stayed at $5 \times 10^{-4}$ per residue per generation (corresponding to a per protein per generation mutation rate of $\mu=10^{-2}$), and at each generation all proteins that folded to the target native structure with $\dgf \le 0$ reproduced with equal probability.  We then evolved each of these equilibrated populations for a further 5,000 generations to collect data. We combined the data for all the folded proteins in the final populations of all the replicates to calculate the average number of mutations \mavg\ after $T$ generations, the corresponding index of dispersion \iod, and the distribution of stabilities shown in Fig. \ref{fig:lattice}.  If we instead simply randomly chose a single folded protein from the final population of each replicate, we obtained results that were identical within the precision shown in Fig. \ref{fig:lattice}.  We emphasize that \mavg\ and \iod\ were computed by keeping track of the actual number of mutations that had occurred during the evolutionary history of each protein, not simply by counting the number of amino acid differences between the ancestral and final sequences (the two quantities may differ if a single site undergoes multiple mutations, as discussed in more detail in later sections).

To generate the data for $\Npop\mu \gg 1$, we used the same procedure but with $N = 10^5$ and only performed 10 replicates.  We again computed the statistics shown in Fig. \ref{fig:lattice} by  combining the data for all of the folded proteins in the final populations of all $10$ replicates.  Similar results were obtained if we instead computed \mavg\ and \iod\ over all of the folded proteins in the final population of a single replicate (average values of \mavg\ were identical  while the \iod\ values of 1.03, 0.95, and 0.94 were extremely similar to those shown from top to bottom in Fig. \ref{fig:lattice}).  This outcome is expected since the probability distributions for $\Npop\mu \gg 1$ evolve deterministically.  

\subsection*{Lattice Protein Predictions}
The numerical predictions for the lattice proteins given in Fig. \ref{fig:lattice} were computed by constructing the matrix \W\ described in the first section of RESULTS with a bin size of $b = 0.005$ and truncating the matrix by assuming that no proteins would have stabilities less than -5.0.  For the case when $\Npop\mu \ll 1$, \mavg\ was calculated using Equation \ref{eq:mavgsmall} and \iod\ was calculated using Equation \ref{eq:iodsmallapprox}.  For $\Npop\mu \gg 1$, \mavg\ was calculated using Equation \ref{eq:mavginf} and \iod\ was calculated using Equation \ref{eq:iodinf}.

\section*{Results}
\subsection*{Assumptions and Mathematical Background}
In this section we describe the physical view of protein evolution that motivates our work.  We begin with the basic observations that evolution selects for protein function, and that most proteins must stably fold in order to function \cite{Anfinsen1973}, meaning that protein stability is under evolutionary pressure only insofar as it must be sufficient to allow a protein to fold and function.   In taking this view, we ignore those proteins (estimated at 10\% of prokaryotic and 30\% of  eukaryotic proteins) that are intrinsically  disordered~\cite{Uversky2005}, as well as those rare proteins that are only kinetically stable~\cite{Jaswal2002}. Natural selection for function requires a protein to fold with some minimal stability \dgfmin, since proteins that lack this minimal stability will be unable to reliably adopt their native structure and perform their biochemical task. A protein's extra stability beyond this minimal threshold is quantified as $\dgfextra = \dgf - \dgfmin$, meaning that all functional proteins must have $\dgfextra \le 0$ (more negative values of \dgf\ indicate increased stability). We further assume that as long as $\dgfextra \le 0$, natural selection for protein function is indifferent to the exact amount of extra stability a protein posesses.  This assumption is at odds with the persistent speculation that high stability inherently impairs protein function and so is selected against by evolution~\cite{DePristo2005,Somero1995}.  But the circular argument most commonly advanced to support this speculation --- that the observed marginal stability of natural proteins indicates that higher stability is detrimental to protein function --- has now been contradicted both by experiments that have dramatically increased protein stability without sacrificing function~\cite{Serrano1993,Giver1998,vandenBurg1998,Zhao1999} and by demonstrations that marginal stability is a simple consequence of the fact that most mutations are destabilizing~\cite[, as well as the current work]{Taverna2002,Arnold2001}. There is a possibility, however, that certain regulatory proteins must be marginally stable to faciliate rapid degradation~\cite{Huntzicker2006}.  To summarize, current biochemical evidence supports our assumption that (with certain well-defined exceptions) the only requirement imposed on protein stability by natural selection for protein function is that stability must meet or surpass some minimal threshold (a protein must have $\dgfextra \le 0$).

\begin{figure}
\begin{psfrags}
\psfrag{dGf}[tc][tc]{\dgfmut (kcal/mol)}
\psfrag{fraction of mutants}[bc][bc]{fraction of mutants}
\psfrag{stability cutoff}[bc][bc]{\dgfmin}
\psfrag{folded proteins}[rb][rb]{folded}
\psfrag{unfolded proteins}[lb][lb]{unfolded}
\psfrag{wildtype dGf = -7.5}[bc][bc]{\dgfwt}
\psfrag{-10}[cc][cc]{\small -10}
\psfrag{0}[cc][cc]{\small 0}
\psfrag{0.1}[cc][cc]{\small 0.1}
\psfrag{0.2}[cc][cc]{\small 0.2}
\psfrag{0.3}[cc][cc]{\small 0.3}
\psfrag{-5}[cc][cc]{\small -5}
\psfrag{5}[cc][cc]{\small 5}
\centerline{\includegraphics[width=8.7cm]{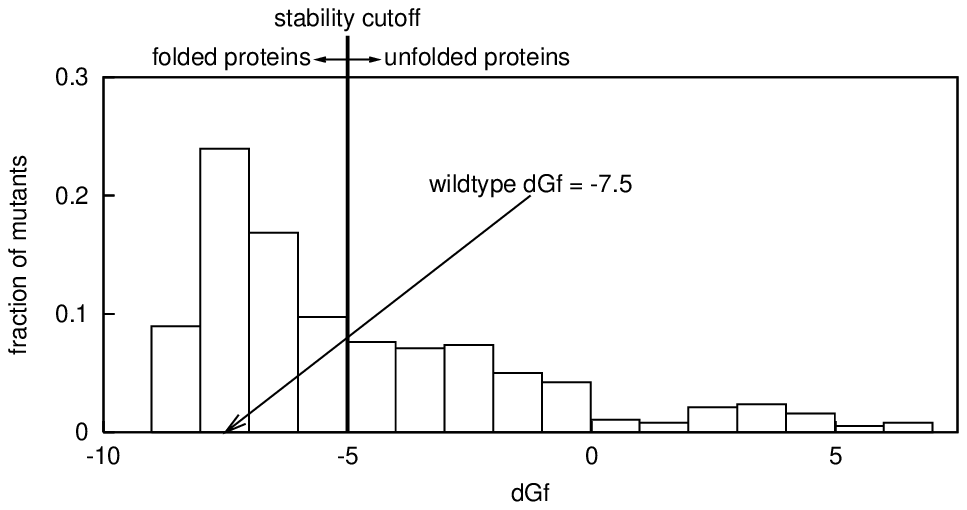}}
\end{psfrags}
\caption{\label{fig:view}A thermodynamic view of protein evolution. A mutant protein stably folds if and only if it possesses some minimal stability, \dgfmin\ (in this case -5 kcal/mol). The stability of the wildtype protein is $\dgfwt = -7.5$ kcal/mol, meaning that it has $\dgfextra = -2.5$ kcal/mol of extra stability.  The bars show the distribution of \ddg\ values for mutations.  Those mutants with $\dgfextra + \ddg \le 0$ still stably fold, while all other mutants do not fold and so are culled by natural selection. The probability that a mutation will be neutral with respect to stable folding is simply the fraction of the distribution that lies to the left of the threshold.  The data in this figure are hypothetical.}
\end{figure}

A mutation to a protein changes its stability by an amount \ddg, and experimental measurements of \ddg\ values have shown that most mutations are destabilizing (have $\ddg > 0$)~\cite{Pakula1986,Godoy-Ruiz2004,Matthews1993,Kumar2006}. A mutation is neutral with respect to selection for stability if $\ddg + \dgfextra \le 0$ since the mutant protein still satisfies the minimal stability threshold; otherwise the mutant does not stably fold and is culled by natural selection. Of course, mutations can also have specific effects on protein function (such as altering an enzyme's activity), but experiments have shown that such mutations are rare compared to the large number of mutations that affect stability~\cite{Shortle1985,Pakula1986,Loeb1989,Bloom2006}. Mutations can also have effects unrelated to the functioning of the individual protein molecule: they can affect its propensity to aggregate~\cite{Chiti2000}, alter its codon usage~\cite{Akashi2003},  change its mRNA stability~\cite{Chamary2005}, affect the efficiency or accuracy of translation~\cite{Akashi2003,Rocha2004}, or change the fraction of mistranslated proteins that fold~\cite{Drummond2005}.  These higher-level effects are probably most apparent in the evolution of highly expressed proteins~\cite{Drummond2005,Pal2001}.  However, here we ignore such effects and assume that the evolutionary impact of a mutation is mostly determined by its effect on protein stability (an assumption in agreement with a recent bioinformatics analysis by \citeN{Sanchez2006}).  The view we present therefore describes the impact of a mutation solely by its \ddg\ value and the \dgfextra\ of the wildtype protein, and is summarized graphically in Fig. \ref{fig:view}  We have previously used a similar view to successfully describe experimental protein mutagenesis results~\cite{Bloom2005,Bloom2006}.

To use the view of Fig. \ref{fig:view} to construct a useful description of neutral protein evolution, we make one major assumption: that the overall distribution of \ddg\ values for random mutations stays roughly constant as the protein sequence evolves.  Actually, this assumption is stronger than is strictly needed for the mathematical theory presented below --- the theory can be developed simply by assuming that all proteins with the same \dgf\  have the same distribution of \ddg\ values (in this case the matrix elements \Wij\ introduced below depend on $j$ in addition to the difference $i-j$).  However, we make the stronger assumption that the \ddg\ distribution remains constant during sequence evolution, since we believe that this assumption is consistent with existing evidence.  We emphasize that this assumption does not imply that we are arguing that the \ddg\ distribution is identical for every possible protein sequence.  Clearly, for any given structure there is a most stable sequence (with all \ddg\ values positive), a least stable sequence (with all \ddg\ values negative), and a vast range of sequences in between.  However, most of these sequences fall within a stability range that is never populated by evolution, since simulations \cite{Taverna2002} and experiments \cite{Keefe2001,Davidson1995} clearly show that the vast majority of protein sequences do not stably fold into any structure (meaning the least stable folded protein is still far more stable than the typical random sequence).  Among the subset of sequences that do stably fold, the simple statistical reality that marginally stable sequences are far more abundant than highly stable sequences causes evolution to further confine itself mostly to sequences with stabilities far less than that of the most stable sequence \cite[, as well as the current work]{Taverna2002,Arnold2001}.  This fact is amply demonstrated by engineering experiments that have greatly increased the stability of natural proteins without sacrificing any of their functional properties \cite{Serrano1993,Giver1998,vandenBurg1998,Zhao1999}.  Therefore, although the distribution of \ddg\ values certainly varies widely among all sequences, it is still reasonable to assume that it is relatively constant among those sequences visited by natural evolution.  This assumption of a constant \ddg\ distribution among evolved sequences is explicitly supported by simulations~\cite[, as well as the current work]{Bloom2005,Bloom2006,Broglia1999,Wilke2005}, and is consistent with the observation that the number of neighbors on a protein's neutral network is approximately determined by its stability~\cite{Bornberg-Bauer1999,Xia2004b}.  Furthermore, protein mutagenesis experiments indicate that the \ddg\ values for random mutations are usually additive~\cite{Serrano1993,Wells1990}, meaning that any given mutation to a protein of length $L$ will alter only $\approx 1/L$ of the other \ddg\ values, leaving the \ddg\ distribution mostly unchanged. Finally, the assumption of a constant \ddg\ distribution has been shown to explain the experimentally observed exponential decline in the fraction of functional proteins with increasing numbers of mutations~\cite{Bloom2005}.  However, we acknowledge that at present the assumption of a roughly constant \ddg\ distribution among neutrally evolving proteins can be verified only for lattice proteins --- for real proteins the most we can say is that it is consistent with existing experimental evidence.

We begin our mathematical treatment by conceptually dividing the continuous variable of protein stability into small discrete bins of width $b$.  This discretization of stability allows us to treat mutations as moving a protein from one bin to another --- the bins can be made arbitrarily small to eliminate any numerical effects of the binning.  The stability of each folded protein in the evolving population (the folded proteins are all those with $\dgfextra \le 0$) can be described by specifying its stability bin.  Specifically, a protein is in bin $i$ if it has \dgfextra\ between $\left(1-i\right)b$ and $-ib$, where $i = 1, 2, \ldots$. Let \Wij\ be the probability that a random mutation has a \ddg\ value such that it moves a protein's stability from bin $j$ to bin $i$, where $i$ and $j$ both are in the range $1, 2, \ldots$.  Then \Wij\ is easily computed as the fraction of  \ddg\ values between $b\left(j - i - 1\right)$ and $b\left(j - i\right)$.  Since \Wij\ only describes transitions between folded proteins, and since we have assumed that a protein's mutational tolerance is determined by its stability, then the fraction of folded mutants (neutrality) of a protein in bin $j$ is $\nuj = \sum\limits_{i} \Wij$.  Clearly, more stable proteins will have larger values of \nuj.

In the next two sections, we will use the matrix \W\ with elements \Wij\ to calculate the distribution of stabilities in an evolving protein population of constant size \Npop, the mean number of mutations \mavg\ after $T$ generations, the corresponding index of dispersion $\iod = \frac{\msqavg - \mavg^2}{\mavg}$, and the average fraction of mutations \nuavg\ that do not destabilize the proteins past the minimal stability threshold.  We assume that \W\ is computed from the distribution of \ddg\ values for all random single amino-acid mutations, although in principle it could be for any type of mutation.  We also assume that the per-protein-per-generation mutation rate $\mu$ is small, so that at each generation a protein undergoes at most one mutation.  Our  calculations at first follow, and then extend the theoretical treatment by \citeN{vanNimwegen1999} of evolution on a neutral network.  In particular, we follow their lead in separately treating the two limiting cases where the product $\Npop\mu$ of the population size and mutation rate is $\ll 1$ and $\gg 1$. We emphasize that all of the equations derived in the next two sections depend only on the mutation rate $\mu$, the number of generations $T$, and the matrix \W\ which can be computed from the single-mutant \ddg\ values.  The population size $N$ determines the applicable limiting case, but otherwise drops out of all final results.

\subsection*{Limit when $\mathbf{\Npop\mu \ll 1}$}
When $\Npop\mu \ll 1$, the evolving population is usually clonal, since each mutation is either lost or goes to fixation before the next mutation occurs.  If a mutation destabilizes a protein in the population beyond the stability cutoff, then it is immediately culled by natural selection.  If a mutation does not destabilize a protein beyond the stability cutoff, it will be lost to genetic drift with probability $\frac{N-1}{N}$ and go to fixation with probability $1/\Npop$~\cite{Kimura1983}.  Since mutations occur rarely ($\Npop\mu \ll 1$), the loss or fixation of the mutant will occur before the next mutant appears in the population.  The entire population therefore moves as one entity along its neutral network.  The population can thus be described by the column vector $\p\left(t\right)$, with element $p_i\left(t\right)$ giving the probability that the population is in stability bin $i$ at time $t$.

If the population is initially in stability bin $j$, at each generation there is a probability $\Npop\mu\Wij$ that a protein experiences a mutation that changes its stability to bin $i$, and if such a mutation occurs, then there is a probability of $1/N$ that it is eventually fixed in the population.  Therefore, at each generation there is a probability $\mu\Wij$ that the population experiences a mutation that eventually causes it to move from stability bin $j$ to bin $i$.  If we define the matrix \V\ so that the diagonal elements are given by $\Vii = \nui$ and all other elements are zero, then \p\ evolves according to
\begin{eqnarray}
\label{eq:pevolution}
\p\left(t + 1\right) &=& \left(\I - \mu\V + \mu\W\right) \p\left(t\right)
\end{eqnarray}
where \I\ is the identity matrix.  Note that this equation treats lethal mutations (those that destabilize a protein beyond the cutoff) as immediately being lost to natural selection and so leaving the population in its original stability bin (hence the population accumulates a mutation with probability $\mu\V$ rather than probability $\mu$).  Equation \ref{eq:pevolution} describes a Markov process with the non-negative, irreducible, and acyclic transition matrix $\A = \I - \mu\V + \mu\W$, and so \p\ approaches the unique stationary distribution \pavg\ satisfying
\begin{equation}
\label{eq:pavg}
0 = \left(\V - \W\right)\pavg.
\end{equation}
This equation gives the expected distribution of protein stabilities solely in terms of the single-mutant \ddg\ values.

We now calculate the average number of mutations \mavgsmall\ that accumulate in an equilibrated population after $T$ generations and the corresponding index of dispersion \iodsmall.  We emphasize that \mavgsmall\ represents the average number of accumulated mutations during the course of the evolutionary process.  When the number of accumulated mutations $m$ is small compared to the length of the protein sequence $L$ ($m \ll L$), then $m$ is just equal to the number of residues differing from those in the parent protein sequence (the Hamming distance).  However, when $m$ becomes substantial relative to $L$, $m$ becomes larger than the Hamming distance since some sites will undergo multiple mutations~\cite{Jukes1969}.  In this case it is necessary to use a substitution model to infer $m$ from the observed Hamming distance.  In the treatment that follows, we calculate the expected value of $m$; application of these formulae to actual protein sequences requires use of one of the well-established statistical techniques for inferring $m$ from the Hamming distance~\cite{Jukes1969,Goldman1994}.  We begin the calculation of \mavgsmall\ by defining \pmt\ to be the column vector with element $i$ giving the probability that at time $t$ the population has accumulated $m$ mutations and is in stability bin $i$.  The time evolution of \pmt is given by
\begin{equation}
\pmtplus = \left(\I - \mu\V\right)\pmt + \mu\W\pmminust.
\label{eq:pmt}
\end{equation}
The $k$th moment of the number of mutations at time $t$ is
\begin{equation}
\label{eq:mkt}
\mkt = \e \sum\limits_m m^k\pmt,
\end{equation}
where $\e = \left(1, \ldots, 1\right)$ is the unit row vector.  We can write a recursive equation for \monet\ in the long-time limit (steady state) by multiplying both sides of Equation \ref{eq:pmt} by $m$, summing over $m$, and left multiplying by \e\ to obtain
\begin{eqnarray}
\label{eq:mavgsmallrecursion}
\monetplus &=& \e \left(\I - \mu\V\right)\sum\limits_m m\pmt + \mu\e\W \sum_m m \pmminust \nonumber \\ 
&=& \e\A \sum\limits_m m\pmt + \mu\e\W\pavg \nonumber  \\
 &=& \monet + \mu \nusmall,
\end{eqnarray}
where we have used the property  $\e \A = \e$, noted that in the long-time limit $\sum\limits_m \pmt = \pavg$ and $\sum\limits_m m\pmminust = \sum\limits_m \left[\left(m-1\right)\pmminust + \pmminust\right] = \sum\limits_m m\pmt + \pavg$, and defined the average neutrality as $\nusmall = \e\W\pavg = \e\V\pavg$. Summing the recursion yields the steady-state value for the number of accumulated mutations,
\begin{equation}
\label{eq:mavgsmall}
\mavgsmall = T\mu\nusmall.
\end{equation}

To calculate the index of dispersion $\iodsmall = \frac{\msqavgsmall - \mavgsmall^2}{\mavgsmall}$, we need to find the second moment \msqavgsmall. In a fashion analogous to the construction of Equation \ref{eq:mavgsmallrecursion}, we can write a recursive expression for the long-time limit of \msqavgsmall\ as
\begin{eqnarray}
\label{eq:msqavgsmallrecursion}
\mtwotplus 
&=& \e \left(\I-\mu\V\right)\sum\limits_m m^2\pmt + \mu\e\W \sum_m m^2 \pmminust\ \nonumber \\
&=& \e\A \sum\limits_m m^2\pmt  + 2\mu\e\W\sum\limits_m m\pmt + \mu\e\W\pavg \nonumber \\
&=& \mtwot + 2\mu\e\W \left[ \A\sum\limits_m m\pmtminus + \mu\W\pavg \right]  + \mu\nusmall \nonumber \\
&=& \mtwot + 2\mu^2\e\W\sum_{\tau=0}^{t-1}\A^{\tau}\W\pavg + \mu\nusmall
\end{eqnarray}
where we have used the property (implicit in Equation \ref{eq:mavgsmallrecursion}) that in the long-time limit, $\sum\limits_m m\pmt = \A\sum\limits_m m\pmtminus + \mu\W\pavg$. Summing the recursion yields the following value for the long-time limit,
\begin{eqnarray}
\label{eq:msqavgsmall}
\msqavgsmall 
&=& T\mu\nusmall + 2\mu^2 \e\W\sum\limits_{t=0}^{T-1}\sum\limits_{\tau=0}^{t-1}\A^{\tau}\W\pavg \nonumber \\
&=& T\mu\nusmall + 2\mu^2\e\W \sum_{t=1}^{T}\left(T - t\right)\A^{t-1} \W\pavg \nonumber \\ 
&=& T\mu\nusmall + T\left(T - 1\right)\mu^2\nusmall^2 + 2\mu^2\e\W\sum_{t=1}^{T}\left(T - t\right)\left(\A^{t-1} - \Q\right)\W\pavg,
\end{eqnarray}
where we have made the substitution $\e\W\Q\W\pavg = \nusmall^2$ and noted that $\lim_{t\to\infty} \A^t = \Q = \left(\pavg, \ldots, \pavg\right)$ since \A\ is an irreducible, aperiodic, stochastic matrix~\cite{Ewens2005}.  This yields a value for the index of dispersion in the long-time limit of
\begin{equation}
\label{eq:iodsmallexact0}
\iodsmall = 1 - \mu\nusmall + \frac{2\mu}{\nusmall}\e\W\sum_{t=1}^{T}\left(1 - \frac{t}{T}\right)\left(\A^{t-1} - \Q\right) \W\pavg.
\end{equation}
The above equation is consistent with the generic equation for the index of dispersion given by \citeN{Cutler2000} and \citeN{Cutler2000b}, where we now give concrete expressions for the variables $\rho$ and $h\left(t\right)$ in Cutler's formula in terms of measureable quantitites, namely $\rho = \mu\nusmall$ and $h\left(t\right) = \frac{\mu}{\nusmall}\e\W\A^{t-1}\W\pavg$.

We can further simplify Equation \ref{eq:iodsmallexact0} by performing spectral decompositions of \A\ and \Q.  Let $\lambda_1, \ldots, \lambda_K$ be the eigenvalues of $\V - \W$, and let $\rvec_1, \ldots, \rvec_K$ and $\lvec_1, \ldots, \lvec_K$ be the corresponding right and left eigenvectors, normalized so that $\lvec_i\rvec_j = 1$ if $i = j$ and 0 otherwise. These eigenvectors are also eigenvectors of the irreducible, acyclic, stochastic \A, and the corresponding eigenvalues are $1 - \mu\lambda_1, \ldots, 1 - \mu\lambda_K$, with Perron-Frobenius theorems guaranteeing that one eigenvalue (chosen here to be $1-\mu\lambda_1$) is equal to one and all other eigenvalues have absolute values less than one.  Then $\rvec_1$ and $\lvec_1$ are right and left eigenvectors of \Q\ with eigenvalue $1$ (i.e. $\rvec_1 = \pavg$ and $\lvec_1 = \e$), and all other eigenvalues of \Q\ are zero.  The spectral decompositions are therefore $\Q = \rvec_1\lvec_1$ and $\A = \rvec_1\lvec_1 + \sum\limits_{i=2}^K \left(1 - \mu\lambda_i\right)\rvec_i\lvec_i$.  Inserting these spectral decompositions into Equation \ref{eq:iodsmallexact0}, we find for the index of dispersion a value of
\begin{eqnarray}
\label{eq:iodsmallexact}
\iodsmall &=& 1 - \mu\nusmall + \frac{2\mu}{\nusmall}\e\W
\sum_{t=1}^{T}\left(1 - \frac{t}{T}\right) 
\sum_{i=2}^{K}\left(1 - \mu\lambda_i\right)^{t-1}\rvec_i\lvec_i\W\pavg,
\end{eqnarray}
since $\A^t = \rvec_1\lvec_1 + \sum\limits_{i=2}^K \left(1 - \mu\lambda_i\right)^t \rvec_i\lvec_i$ \cite{Ewens2005}.  In the limit of large $T$ and small $\mu$, the value of \iodsmall\ given by the above equation approaches the value
\begin{eqnarray}
\label{eq:iodsmallapprox}
\iodsmall &\approx& 
1 + \frac{2\mu}{\nusmall}\e\W\sum\limits_{t=1}^{T}\sum\limits_{i=2}^{K} \left(1-\mu\lambda_i\right)^{t-1}\rvec_i\lvec_i\W\pavg \nonumber \\
&\approx& 1 + \frac{2}{\nusmall}\e\W\sum_{i=2}^{K}\lambda_i^{-1}\rvec_i\lvec_i\W\pavg,
\end{eqnarray}
where the $\mu\nusmall$ term drops out because $\mu$ is small and the $\sum\limits_{t=1}^{T}\frac{t}{T}\left(1-\mu\lambda_i\right)^{t-1}$ term drops out because $T$ is large and $\left|1-\mu\lambda_i\right| < 1$.  This equation shows that \iodsmall\ approaches a constant value independent of $T$ and $\mu$.  Although we could not prove that value of \iodsmall\ given by Equation \ref{eq:iodsmallapprox} is necessarily greater than one (since some of the eigenvalues $\lambda_i$ could be complex), in all of our simulations we observed $\iodsmall > 1$, suggesting that when $\Npop\mu \ll 1$, fluctuations in protein stability tend to overdisperse the molecular clock.  

\subsection*{Limit when $\mathbf{\Npop\mu \gg 1}$}
When $\Npop\mu \gg 1$, the population is spread across many nodes of the neutral network rather than converged on a single sequence~\cite{vanNimwegen1999}.  In this limit, we treat the evolutionary dynamics of the population deterministically (i.e., we assume an infinite population size), and describe the distribution of stabilities in the population by the column vector $\x\left(t\right)$, with element $x_i\left(t\right)$ giving the fraction of proteins in the population at time $t$ that have stabilities in bin $i$.  At generation $t$, the fraction of mutated proteins that continue to fold is $\nut = \e\W\x\left(t\right)$.  These folded proteins reproduce, and in order to maintain a constant population size, this reproduction must balance the removal of proteins by death, meaning that each folded sequence must produce an average of $\ratet = \left[1 - \mu\left(1- \nut\right)\right]^{-1}$ offspring.  The population therefore evolves according to
\begin{equation}
\label{eq:xinft}
\x\left(t + 1\right) = \ratet\left[\left(1 - \mu\right)\I + \mu\W\right]\x\left(t\right).
\end{equation}
After the population has evolved for a sufficient period of time, \x\ approaches an equilibrium distribution of \xinf. The corresponding equilibrium neutrality is $\nuinf = \e\W\xinf$, and the equilibrium reproduction rate is $\rateinf = \left[1 - \mu\left(1 - \nuinf\right)\right]^{-1}$, so
\begin{equation}
\label{eq:xinf}
\xinf = \rateinf\left[\left(1 - \mu\right)\I + \mu\W\right]\xinf.
\end{equation}
This equation can be rewritten to show that \xinf\ is the principal eigenvector of \W,
\begin{equation}
\label{eq:xinfeigen}
\nuinf\xinf = \W\xinf.
\end{equation}
We note that \nuinf\ approximates the asymptotic neutrality for the decline in the fraction of folded proteins upon random mutagenesis~\cite{Bloom2005,Wilke2005}.

We now determine the average number of accumulated mutations \mavginf\ and the corresponding index of dispersion \iodinf\ by treating the forward evolutionary process.  As described in the text immediately prior to Equation \ref{eq:pmt}, our calculations describe the actual number of mutations accumulated during the evolutionary process, which may differ from the number of sequence differences relative to the ancestor if a single site undergoes multiple mutations.  When $\Npop\mu \gg 1$, it is not \textit{a priori} obvious that the average number of mutations present in the population is equivalent to number of fixed substitutions along the line of descent. Therefore, in the APPENDIX, we show that identical results are obtained by tracing a randomly chosen protein backwards in time along its ancestor distribution, proving the treatment we give below is mathematically equivalent to treating the time-reversed process.  We define \xmt\ as the column vector with element $i$ giving the fraction of the population at time $t$ that has accumulated $m$ mutations and is in stability bin $i$.  Once the population has reached the equilibrium distribution of stabilities, the time evolution of \xmt\ is
\begin{eqnarray}
\label{eq:xinfevolution}
\xmtplus &=& \rateinf\left(1 - \mu\right)\xmt + 
\rateinf\mu\W\xmminust.
\end{eqnarray} 
The recursion can be solved to obtain
\begin{equation}
\label{eq:xinfexplicit}
\xmt = \rateinf^t\sum\limits_{\kappa=0}^t\binom{t}{\kappa}\left(1 - \mu\right)^{t-\kappa}\mu^{\kappa}\W^{\kappa}\x\left(m - \kappa, 0\right),
\end{equation}
as can be verified by direct substitution.  Since we are assuming the population has equilibrated at time 0 and no mutations have accumulated at that time, $\x\left(m, 0\right)$ is \xinf\ for $m = 0$ and 0 otherwise. Furthermore, \xinf\ satisfies Equation \ref{eq:xinfeigen}, so multiplying Equation \ref{eq:xinfexplicit} by \e\ yields
\begin{equation}
x\left(m,t\right) = \binom{t}{m}\rateinf^t\left(1-\mu\right)^{t-m}\left(\mu\nuinf\right)^m,
\label{eq:xfrac}
\end{equation}
where $x\left(m,t\right) = \e\x\left(m,t\right)$ gives the fraction of the population that has accumulated $m$ mutations after $t$ generations. The average number of accumulated mutations after $T$ generations is the mean of this binomial distribution,
\begin{equation}
\mavginf = \frac{T\mu\nuinf}{1 - \mu\left(1 - \nuinf\right)}.
\label{eq:mavginf}
\end{equation}
Using the well known result for the variance of the binomial distribution, we find that the index of dispersion is
\begin{eqnarray}
\label{eq:iodinf}
\iodinf &=& 
1 - \frac{\mu\nuinf}{1 - \mu\left(1 - \nuinf\right)}.
\end{eqnarray}
It is important to reiterate that the above equation was derived under the assumption that there is at most one mutation per sequence per generation.  For realistic distributions of mutations (i.e. Poisson), this means that $\mu \ll 1$.  In this regime, \iodinf\ is close to one.

\subsection*{Lattice Protein Simulations}
\begin{figure}
\centerline{\includegraphics[width=18cm]{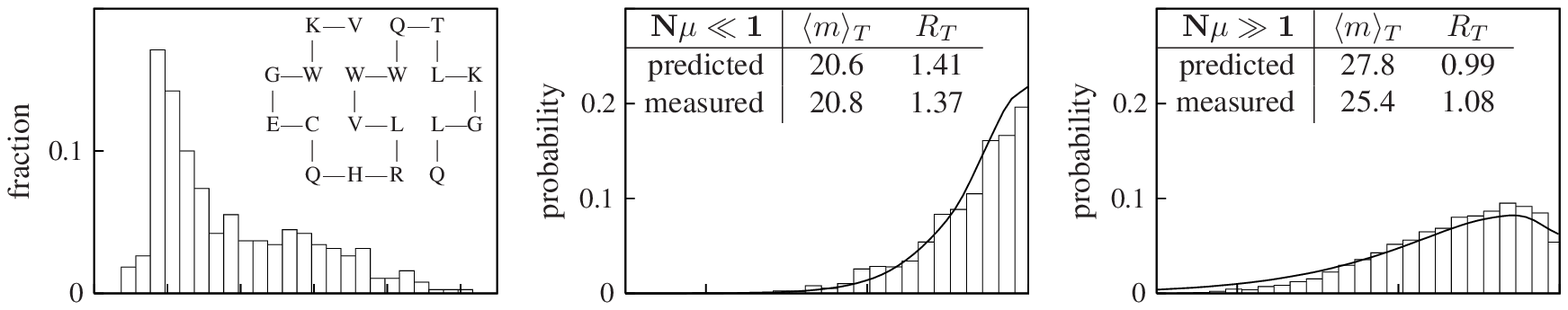}}
\centerline{\includegraphics[width=18cm]{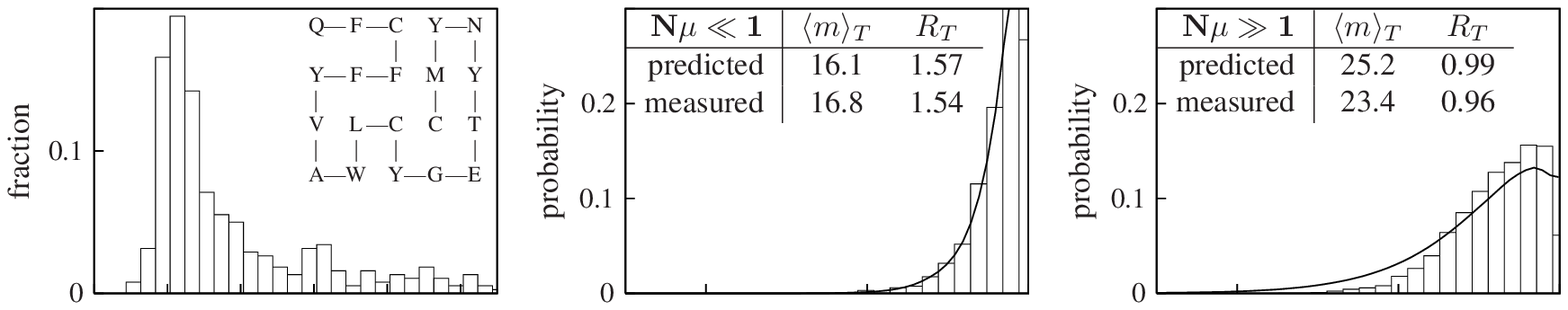}}
\centerline{\includegraphics[width=18cm]{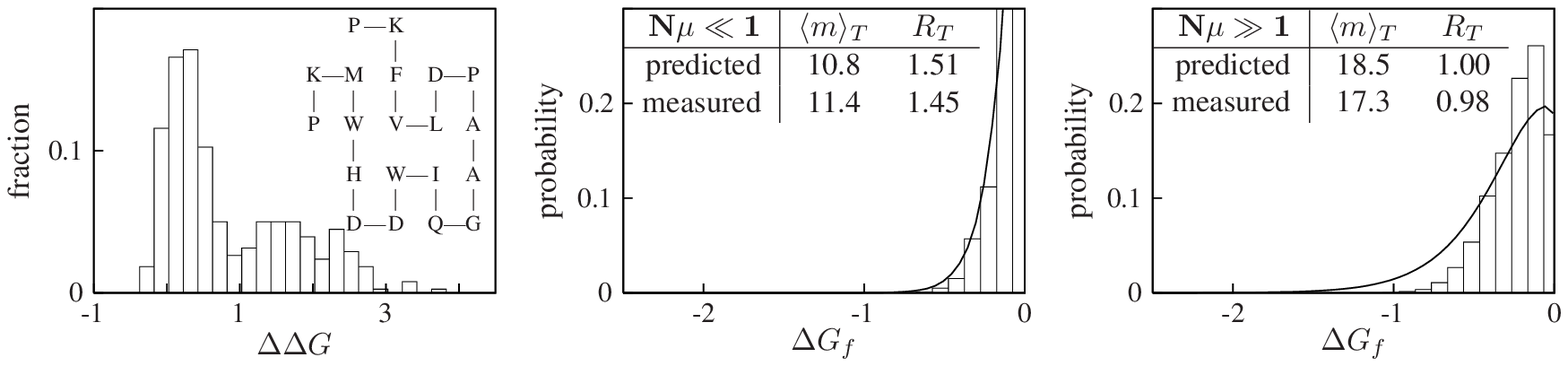}}
\caption{\label{fig:lattice}The theory gives accurate predictions for the evolution of model lattice proteins.  Each row of panels corresponds to a different lattice protein.  The graphs at left show the starting protein and the distribution of \ddg\ values for all point mutations.  The graphs in the middle and right show the predicted (lines) and measured (boxes) distributions of stabilities among the evolved proteins.  The tables embedded in the graphs show the predicted and measured values for the average number of mutations (\mavg) and the index of dispersion (\iod) after 5,000 generations of neutral evolution.  The center graphs are for a population size of  $N = 10$, and the graphs at the right are for $N = 10^{5}$.  In both cases, the per protein per generation mutation rate is $\mu = 0.01$.  As predicted, the evolving population with $N\mu \gg 1$ evolved mutational robustness that is manifested by increased protein stability.  This additional mutational robustness accelerated the rate of sequence evolution.}
\end{figure}

We tested our theory's predictions on the evolutionary dynamics of lattice proteins.  Lattice proteins are simple protein models that are useful tools for studying protein folding and evolution~\cite{Chan2002}. Our lattice proteins were chains of $20$ amino acids that folded on a two-dimensional lattice.  The energy of a lattice protein conformation was equal to the sum of the pairwise interactions between non-bonded amino acids~\cite{Miyazawa1985}.  Each lattice protein has 41,889,578 possible conformations, and by summing over all of these conformations we could exactly determine the partition sum and calculate \dgf.  We set a minimal stability threshold for the lattice proteins of $\dgfmin = 0$, meaning that we considered all proteins that folded to the target structure with $\dgf \le 0$ to be folded and functional, while all proteins with $\dgf > 0$ were considered to be nonfunctional.  We note that this stability threshold is equivalent to requiring a lattice protein to spend at least half of its time in the target native structure at equilibrium.  We began by generating lattice proteins that stably folded to each of the three different structures shown in Figure \ref{fig:lattice}.  For each of these three proteins, we determined the distribution of \ddg\ values for all 380 single mutations (these distributions are shown in Figure \ref{fig:lattice}).  These distributions were used to construct the matrix \W\ and to predict the equilibrium distribution of stabilities, the average number of mutations, and the indices of dispersion for both the $\Npop\mu \ll 1$ and the $\Npop\mu \gg 1$ cases, using the equations presented in the preceding sections. 

To test the accuracy of these predictions, we then simulated evolving populations of the lattice proteins with a standard evolutionary algorithm using Wright-Fisher sampling.  Briefly, the populations were held at a constant size of either $\Npop = 10$ or $\Npop = 10^{5}$.  At each generation, a new population was created by choosing parents with equal probability from all folded proteins in the previous generation's population, and copying these parents into the new population with a mutation rate of $5\times 10^{-4}$ mutations per residue per generation.  Since the proteins have a length of $20$ amino acids, this mutation rate corresponds to a per-protein-per-generation mutation rate of $\mu = 10^{-2}$.  Therefore, the product $\Npop\mu$ is either $0.1$ or $10^{3}$, corresponding to $\Npop\mu \ll 1$ or $\Npop\mu \gg 1$, respectively.  We emphasize that the lattice protein evolutionary algorithm is the same for both population sizes.  When $\Npop = 10$ the population naturally follows dynamics approximating those presented for $\Npop\mu \ll 1$, while when $\Npop = 10^{5}$ it naturally follows dynamics approximating those presented for $\Npop\mu \gg 1$ (as evidenced by the excellent agreement of the predictions with the simulations).  For $\Npop = 10$, we performed 1,000 replicates for each different structure.  For $\Npop = 10^{5}$, computational constraints limited us to 10 replicates for each structure (however the evolutionary dynamics are nearly deterministic in this case, so all replicates yielded similar results).  We note that during the simulations we recorded the number of mutations that actually accumulated rather than simply computing the number of differences (Hamming distance) from the original sequence.  

Figure \ref{fig:lattice} shows the theoretical predictions and simulation results for each of the three structures.  The theoretical predictions are in good agreement with the simulation results.  Figure \ref{fig:lattice} clearly shows that when $\Npop\mu \gg 1$, the proteins tend to be more stable than when $\Npop\mu \ll 1$.  This extra stability is a biophysical manifestation of the neutrally evolved mutational robustness predicted by \citeN{vanNimwegen1999}. This increase in stability leads to a substantial increase number of accumulated mutations.  In accordance with the theoretical predictions, when $\Npop\mu \ll 1$ the index of dispersion is elevated above one by fluctuations in protein stability.  Another clear results from the simulations is that proteins of different structure show markedly different distributions of stabilities and rates of sequence evolution due to the differences in their \ddg\ distributions.  Overall, the simulations offer strong support for the validity of the theoretical predictions in the preceding sections.

\section*{Discussion}
We have presented that a theory that offers quantitative predictions about the distribution of stabilities, the average number of fixed mutations, and the index of dispersion for an evolving protein population in terms of the \ddg\ values for individual mutations.  We have demonstrated that these predictions are accurate for simple lattice proteins, and have used existing biophysical evidence to argue that the basic theoretical assumptions should also be accurate for real proteins.  In this section, we give qualitative interpretations of the mathematical results and discuss their implications for our understanding of protein evolution. 

One major result is to show that the effects of protein structure on the rate of sequence evolution can be quantitatively cast in terms of the \ddg\ values for single mutations.  Numerous lattice protein simulations have shown that protein structure can dramatically affect the rate of sequence evolution, since structures that are more ``designable'' (encoded by more sequences) can evolve their sequences more rapidly (as can be seen in Fig. \ref{fig:lattice} of this work) \cite{Govindarajan1997,Bornberg-Bauer1999,Tiana2000,Xia2004b,Li1996,Chan2002,Wingreen2004}.  Unfortunately,  these simulations typically measure structural designability by enumerating a large number of lattice protein sequences, meaning that their findings cannot be extended to real proteins for which such extensive enumeration is impossible.  However, recent theoretical work by \citeN{England2003a} has made progress in connecting designability to observable structural properties, and a bioinformatics analysis based on this theoretical measure of designability indicates that structure indeed influences the evolutionary rate of real proteins \cite{Bloom2006b}.  Our work provides a way to quantitatively relate the structural influences on protein evolution to experimentally measureable \ddg\ values, opening the door to further connecting structural designability and sequence evolution to laboratory stability measurements.  Although thousands of \ddg\ values have been measured experimentally \cite{Kumar2006},
at present there are no large sets of measurements for truly random mutations to a single protein.  When such sets of measurements become available, it should be possible to use them in conjunction with the theory that we have presented to predict the neutralities of real proteins with different structures.

A second important result is to show that protein evolutionary dynamics can depend on the product of population size and mutation rate, $\Npop\mu$.  When $\Npop\mu \gg 1$, the evolving protein population is polymorphic in stability and subject to frequent mutations, so the more stable (and thus more mutationally tolerant) proteins produce more folded offspring.  In contrast, when $\Npop\mu \ll 1$, the population is usually monomorphic in stability and so all members of the population are equally likely to produce folded offspring.  The general tendency for populations to neutrally evolve mutational robustness when $\Npop\mu \gg 1$ has previously been treated mathematically by \citeN{vanNimwegen1999}, and a variety of lattice protein simulations have noted the tendency of evolving protein populations to preferentially occupy highly connected neutral network nodes~\cite{Bornberg-Bauer1999,Taverna2002b,Xia2004b}.  Our work shows that for proteins, in the limiting cases when $\Npop\mu \ll 1$ or $\gg 1$, this process can be rigorously described by considering only protein stability, rather than requiring a full analysis of the neutral network (provided, as we have argued is likely to be the case, that the assumption of a roughly constant \ddg\ distribution holds for real proteins as well as it holds for our lattice proteins).  In addition, we prove that the number of accumulated mutations depends on whether $\Npop\mu$ is $\ll 1$ or $\gg 1$.  This finding is at odds with the standard prediction~\cite{Kimura1987} of Kimura's neutral theory that the rate of evolution is independent of population size.  The reason for this discrepancy is that the standard neutral theory fails to account for the possibility that increasing the population size so that $\Npop\mu \gg 1$ can systematically increase the fraction of mutations that are neutral.

A third important contribution of our theory is to use the distribution of \ddg\ values for single mutations to predict the distribution of protein stabilities in an evolving population.  Several researchers have pointed out that evolved proteins will be marginally stable simply because most mutations are destabilizing~\cite{Taverna2002,Arnold2001}; we have described this process quantitatively.  In addition, we have shown how the neutral evolution of mutational robustness when $\Npop\mu \gg 1$ will shift the proteins towards higher stabilities (as shown in Fig. \ref{fig:lattice}), although this increase in stability is limited by the counterbalancing pressure of predominantly destabilizing mutations. The formulae we provide can in principle be combined with experimentally measured \ddg\ values to predict the expected range of stabilities for evolved proteins.

Our work also weds Takahata's concept that fluctuating neutral spaces might overdisperse the molecular clock~\cite{Takahata1987,Cutler2000b,Bastolla2002} to a concrete discription of how protein neutrality fluctuates during evolution.  When $\Npop\mu \ll 1$, fluctuations in protein stability can cause an overdispersion in the number of accumulated substitutions that can be calculated from the single-mutant \ddg\ distribution.  Furthermore, given our assumption of a roughly constant \ddg\ distribution, we show that the index of dispersion will approach a constant value that is independent of time or mutation rate, but will depend on whether $\Npop\mu \ll 1$ or $\gg 1$.  Previous simulations have indicated that overdispersion indeed depends on the population size~\cite{Bastolla2002,Wilke2004}  --- we have explained this dependence by showing that stability-induced overdispersion does not occur when $\Npop\mu \gg 1$ since the population's distribution of stabilities equilibrates as it spreads across many sequences.  Mathematically, the difference in the cases $\Npop\mu \gg 1$ and $\Npop\mu \ll 1$ is that, assuming the \ddg\ distribution remains relatively constant, when the population size is sufficiently large, the distribution of protein stabilities no longer fluctuates in a manner that influences the probability of a substitution (Equation \ref{eq:pmt} contains $\mu\V$ in the first term on the right side, while Equation \ref{eq:xinfevolution} does not).

In summary, we have presented a mathematical theory of how thermodynamics shape neutral protein evolution.  A major strength of our theory is that it makes quantitative predictions using single-mutant \ddg\ values, which can be experimentally measured.  Our work also suggests how neutral and adaptive protein evolution may be coupled through protein thermodynamics.  Protein stability represents an important hidden dimension in the evolution of new protein function, since extra stability that is itself neutral can allow a protein to tolerate mutations that confer new or improved functions~\cite{Bloom2006}. Our theory describes the dynamics of protein stability during neutral evolution --- adaptive protein evolution is superimposed on these stability dynamics, with proteins most likely to acquire beneficial mutations when they are most stable.

\section{Acknowledgements}
We thank Frances H. Arnold for helpful comments and discussion.  J.D.B. is supported by a HHMI pre-doctoral fellowship.  C.O.W. is supported by the National Institutes of Health grant AI 065960.  A.R. is supported by the National Science Foundation grants CCF 0523643 and FIBR 0527023.

\section{Appendix}
Here we calculate the properties of the evolving population when $\Npop\mu \gg 1$ by analyzing the time-reversed process to compute the mean and variation
in the number of mutations in a single randomly chosen protein over time.
We show that the results so obtained are identical to those found in the
main text, where we analyzed the forward-time process to compute the mean and
variation in the number of mutations across the population of evolving
proteins.

When $\Npop\mu \gg 1$, the population
is now never converged to a single sequence, so it is not
\textit{a priori} obvious that the average number of mutations present
in the population is equivalent to the expected number of fixed
substitutions along the line of descent.  In fact, in the limit
of very large population sizes there may not even be a common line of 
descent in relevant time frames, since many new mutations
will occur before any given mutation goes to fixation.  In the
main text we calculated
the average number of mutations \mavginf\ a sequence in the
population has accumulated over the last $T$ generations by treating
the forward evolution of the population.  Here we trace a 
randomly chosen protein in the
population back in time, and show that the average number of substitutions
\savg\ that it has accumulated over the last $T$ generations is equal
to \mavginf.  We also show that indices of dispersion of \mavginf\ and
\savg\ have the same value of \iodinf.

To calculate \savg, we first define
a vector \avec\ giving the ancestor distribution \cite{Hermisson2002}:
element $i$ of 
$\avec\left(T - t\right)$ gives the probability that a randomly chosen
sequence from the population at time $T$ had a predecessor
with stability in bin $i$ at time $T - t$.  The transition probabilities of
$\avec\left(T - t\right)$ when the population is in equilibrium are the discrete time analogue of those
computed by \citeN{Hermisson2002}.  From Equation
\ref{eq:xinfevolution} of the main text, it follows that the fraction of sequences
in bin $i$ at time $t + 1$ that had as their ancestor in the previous
generation a sequence in bin $j$ is 
$\ratet\left[\left(1-\mu\right)\delta_{ij} + \mu\Wij\right]x_j\left(t\right)$.
In order to obtain the probability that a sequence in bin $i$ at time $t+1$
had an ancestor in bin $j$, we have to divide this fraction by the total number of sequences in bin $i$ at time $t+1$.
When the population is at equilibrium,
$\ratet = \rateinf$ and $x_i\left(t+1\right) = x_i\left(t\right) = x_i$ where
$x_i$ is the element from \xinf.  Hence, the probability that a sequence in bin
$i$ had an ancestor in bin $j$ is $\rateinf\left[\left(1-\mu\right)\delta_{ij} + \mu \Hji\right]$,
where we have defined
\begin{equation}
\label{eq:Hji}
\Hji = \Wij x_j/x_i,
\end{equation}
The time evolution of \avec\ is therefore
\begin{equation}
\label{eq:a}
\avec\left(T-t\right) = \rateinf\left[\left(1-\mu\right)\I + \mu\Hm\right]\avec\left(T-t+1\right),
\end{equation}
where the matrix \Hm\ is defined by Equation \ref{eq:Hji}.  
Equation \ref{eq:a} 
can be solved to show that the 
equilibrium value of \avec\ is \ainf\ satisfying
\begin{equation}
\label{eq:ainf}
\nuinf\ainf = \Hm\ainf.
\end{equation}
If we define $\avec\left(s, T - t\right)$ as the vector with element $i$
giving the probability that a randomly chosen sequence at time $T$ had
a predecessor at time $T-t$ in stability bin $i$ and with $s$ substitutions
relative to the sequence at time $T$, then the time evolution for an
equilibrated population is
\begin{equation}
\avec\left(s,T-t-1\right) = \rateinf\left(1-\mu\right)\avec\left(s,T-t\right)
+ \rateinf\mu\Hm\avec\left(s-1,T-t\right).
\label{eq:avecst}
\end{equation}
We can solve Equations \ref{eq:avecst} and \ref{eq:ainf} in a manner analogous to
the forward process to obtain
\begin{equation}
\avec\left(s, T - t\right) = \binom{t}{s}\rateinf^t\left(1-\mu\right)^{t-s}\left(\mu\nuinf\right)^s\ainf.
\label{eq:aexplicit}
\end{equation}
Again defining $a\left(s, T-t\right) = \e\avec\left(s,T-t\right)$ as the 
probability of having accumulated $s$ substitutions as one moves back $t$
generations from time $T$, we obtain the binomial distribution
\begin{equation}
a\left(s,T-t\right) = \binom{t}{s}\rateinf^t\left(1-\mu\right)^{t-s}\left(\mu\nuinf\right)^s.
\label{eq:afrac}
\end{equation}
Comparison of Equation \ref{eq:xfrac} of the main text and Equation 
\ref{eq:afrac} shows that they
are identical.  Therefore, all moments computed from the two distributions
must be equal.  In particular, this proves that $\mavginf = \savg$, and
that the corresponding indices of dispersion have the same
value of \iodinf\ defined by Equation \ref{eq:iodinf} of the main text.
This shows that when $\Npop\mu \gg 1$, we expect equivalent results regardless
of whether we average over the number of mutations in all sequences present
in the population, or randomly choose a single sequence and trace back along
its ancestor distribution.

\end{document}